\documentclass[12pt]{article}
\usepackage{graphicx}
\usepackage{geometry}
\usepackage[english]{babel}
\usepackage{placeins}
\usepackage{amsmath}
\usepackage{verbatim}
\usepackage{braket}
\usepackage{hhline}
\usepackage[hypcap]{caption}
\usepackage{amssymb}
\usepackage[square,numbers,compress]{natbib}
\usepackage{authblk}
\begin{document}


\title{Inverse classical scattering using fractional derivative
}

\author[1]{F. S. Carvalho\footnote{felipe.s.carvalho\_qui@hotmail.com}}
\author[1]{J. P. Braga\footnote{jjppbraga@gmail.com}}
\author[2]{N. H. T. Lemes}
\affil[1]{Departmento de Qu\'imica - ICEx, Universidade Federal de Minas Gerais, 31270-901 - Belo Horizonte, MG, Brazil.}
\affil[2]{Instituto de Qu\'imica, Universidade Federal de Alfenas, 37130-001 - Alfenas, MG, Brazil.}

\maketitle
\clearpage

\begin{abstract}
	The fractional calculus framework will be used to invert the potential energy function from the classical scattering angle, which will be related to Riemann-Liouville fractional integral. Numerical solution of this fractional order problem will be applied to the inverse Rutherford  scattering and to the inverse scattering of Xe--Rn atoms, in which the potential is given by Lennard-Jones function. Proofs of existence will be presented for more clarity and completness of the present work. 
	In the two cases considered, the potential energy function can be retrieved with a desired precision. The present method gives a clear understanding of the inverse fractional problem framework.
\end{abstract}

\textbf{keywords:} Riemann-Liouville fractional integral; classical scattering; inverse problem; numerical solution.

\maketitle 

\clearpage
\section{Introduction}
The fractional calculus area, together with its mathematical interest, has proved to be very important in science. The subject gives new insights into old problems, such as the one to be treated in the present work. For an historical review, theory and applications of the subject the reader is recommended
to \cite{podlubny1998,miller1993,tarasov2013}.\\

Retrieving system information from experimental data is another important subject of research, known as the inverse problem area. This has been explored for several problems in physical-chemistry, such as inverting the heat capacity \cite{JP2014}, 
nuclear magnetic ressonance data \cite{JP2006}, force field potential energy \cite{borgesJP2006} and for the inversion of quantum differential cross section \cite{JP2008}. \\

An overlap between these two subjects, fractional order and inverse problems areas, is to be discussed in the present work, treating the classical Rutherford scattering with additional studies in the Lennard-Jones potential. Retrieving the potential energy function from scattering angle has been discussed before under the inverse problems framework \cite{JP2012}. For the repulsive case, the Firsov approach is one rare example in which the inverse kernel can be obtained analytically. As can be shown, the use of the fractional order methodology provides a straightforward way to obtain the inverse kernel. \\

The Riemann-Liouville fractional integral will be rewritten as a series that is general for any order of integration. 
The series obtained will be applied to the Rutherford scattering and to a potential with repulsive and atractive forces, represented by a Lennard-Jones type potential. In both cases the potential energy function can be obtained with the required precision. An example with four terms in the serie is presented for the Rutherford scattering, providing a simple way to retrive the potential energy function for this repulsive scattering case.

\section{The fractional calculus background}
In this section it will be presented two mathematical concepts in fractional calculus, needed to develop the present work: the Riemann-Liouville fractional integral and the Riemann-Liouville fractional derivate.
The Riemann-Liouville general form for a n-th order integral is expressed as

\begin{equation}\label{eq:RL:int}
J^n f(x) = \frac{1}{\Gamma(n)}\int_{0}^{t} (t-x)^{n-1} f(x)dx
\end{equation}

which has its fractional counterpart,
%

\begin{equation}\label{eq:RL:frac:int}
J^\alpha f(x) = \frac{1}{\Gamma(\alpha)}\int_{0}^{t} (t-x)^{\alpha-1} f(x)dx
\end{equation}

for the non-integer number $\alpha$. Based on reference that gives a more general demonstration for k-Riemann-Liouville fractional integral with a arbritary lower limit\cite{karaca2014}, a similar demonstration is given in this work for clarity and completeness of the paper. To prove that this integral converges for $\alpha>0$ consider the integral over a triangle

\begin{equation}\label{eq:int:Pf}
\int_{0}^{b}\int_{0}^{t} \left| f(x)P(x,t) \right|dxdt
\end{equation}
and that $f(x)$ is a piecewise continuous and integrable in every finite subinterval of $(0,\infty)$ and the function $P(t,x)$ is defined as

\begin{equation}
P\left(x,t\right) = 
\left\lbrace\begin{tabular}{ll}
$\left( t -x  \right)^{\alpha-1}$& \quad $a \le x \le t \le b$\\
$0$              &  \quad $a \le t < x \le b$
\end{tabular}\right.
\end{equation}
with $P(x,t)\ge 0$. Using the Dirichlet's relation,
one obtains

\begin{equation}
\int_{0}^{b}\int_{x}^{b} (t-x)^{\alpha-1}|f(x)|dtdx
\end{equation} 

For $\alpha>0$ there is a singularity at $t=x$. Thus the inferior limit of the integrals will be set as $\epsilon_1$ and $x+\epsilon_2$, in the limit of $\epsilon_1,\epsilon_2\rightarrow0$, as follows

\begin{eqnarray}
\lim\limits_{\epsilon_1,\epsilon_2\rightarrow0}\int_{\epsilon_2}^{b}\int_{x+\epsilon_1}^{b} (t-x)^{\alpha-1}|f(x)|dtdx & = & \lim\limits_{\epsilon_1,\epsilon_2\rightarrow0} \int_{\epsilon_2}^{b} \frac{(b-x)^\alpha - \epsilon_1^\alpha}{\alpha} |f(x)| dx \le \nonumber \\
\lim\limits_{\epsilon_1,\epsilon_2\rightarrow0} \frac{(b-\epsilon_2)^\alpha - \epsilon_1^\alpha}{\alpha} \int_{\epsilon_2}^{b} |f(x)| dx & = &  \frac{b^\alpha}{\alpha}\int_{0}^{b} |f(x)| dx < \infty
\end{eqnarray}

Therefore the integral (\ref{eq:int:Pf}) satisfies the Fubini's theorem
 and the integral 

\begin{equation}
\int_{0}^{t}(t-x)^{\alpha-1}f(x)dx
\end{equation}
converges. \\

The Riemann-Liouville fractional derivative,

\begin{equation}
D^\alpha f(x) = \frac{d^m}{dt^m}(J^{m-\alpha}f(x))(t)
\end{equation}
with $m$ integer positive and $\alpha$ non-integer will be used in the present context. 
This derivative has the important property,

\begin{equation}\label{eq:DJ}
\begin{array}{clcr}
D^\alpha J^{\alpha} g(y) 
&= \frac{d^m}{dt^m}J^{m-\alpha}J^{\alpha} g(y)\\[2mm]
&= \frac{d^m}{dt^m} J^{m}g(y) \\[2mm]
&= g(t)
\end{array}
\end{equation}

as in the calculus of integer order. This property of the Riemman-Liouville derivative will be used to solve the classical Abel's problem. \cite{abel_2012_2}

\section{Scattering angle as a fractional integral}
The classical scattering angle, $\chi$, for a collision between two particles is given by \cite{murrell1989},
$
\chi(E) = \pi - 2\theta(E)$,
%
with

\begin{equation}\label{int:theta}
\theta(E) = b\int_{r_c}^{\infty}\frac{dr}{r^2\sqrt{1 - \frac{E_p(r)}{E} - \frac{b^2}{r^2}}}
\end{equation}
$r_c$ is the turning point
and $b$ is the impact parameter. Defining \cite{carter1999} 
$x=\frac{1}{r}$, $U(x) = E_p(x) + Eb^2x^2$, $dx=\frac{dx}{dU}dU$ and using
%
$E = E_p(r_c) + \frac{Eb^2}{r_c^2}$,
%
one obtains,

\begin{equation}\label{eq:theta:E:0}
\theta(E) = \int_{0}^{E} \frac{b\sqrt{E} \left( \frac{dx(U)}{dU} \right)}{\sqrt{E - U(x)}}dU
\end{equation}


After multiplying and dividing equation (\ref{eq:theta:E:0}) by $\Gamma\left( \frac{1}{2} \right)$ the expression can be written as a Riemann-Liouville fractional integral, 


\begin{equation}
\theta(E) = \left(J^{\frac{1}{2}} \Gamma\left(\frac{1}{2}\right)b\sqrt{E} \left( \frac{dx(U)}{dU} \right)\right)(E)
\end{equation}

Applying $D^{\frac{1}{2}}$ in both sides and using (\ref{eq:DJ}), 

\begin{equation}
\begin{array}{clcr}
\Gamma\left(\frac{1}{2}\right)b\sqrt{E} \left( \frac{dx(U)}{dU} \right) 
&= \frac{d}{dU}\left( \Gamma\left(\frac{1}{2}\right)b\sqrt{E} x(U) \right)\\[2mm] 
&= \left(D^{\frac{1}{2}}\theta(E)\right) \\[2mm]
&= \frac{d}{dU}\left(J^{\frac{1}{2}}\theta(E)\right)
\end{array}
\end{equation}
Therefore,

\begin{equation}\label{eq:xU}
x(U) = \frac{1}{\Gamma\left(\frac{1}{2}\right)b\sqrt{E}}J^{\frac{1}{2}}\theta(E) = \frac{\sqrt{2\mu}}{\Gamma\left(\frac{1}{2}\right)L}J^{\frac{1}{2}}\theta(E)
\end{equation}
given $L=b\sqrt{2\mu E}$ for the angular momentum. The same result can be obtained without the usage of fractional calculus, as in the Abel's original work\cite{abel_2012_2}, however this methodology is less laborious.

\section{Riemann-Liouville integral by Gauss-Mehler quadrature}
The fractional integral, as in equation (\ref{eq:xU}), will be rewritten as a Gauss-Mehler quadrature and transformed into a series solution representation. The Gauss-Mehler quadrature\cite{abramowitz1965} is given by

\begin{equation}
\int_{-1}^{1}\frac{f(x)}{(1-x^2)^\frac{1}{2}}dx \approx \frac{\pi}{n}\sum_{i}^{n}f\left( k \right)
\label{eq8}
\end{equation}
with $k=\cos\left( \frac{(2i-1)\pi}{2n} \right)$. One can use this result for fractional integrals, making the apropriate change of variables,

\begin{equation}
J^\alpha f(x) = \frac{1}{\Gamma(\alpha)}\int_{0}^{t} \frac{f(x)}{\left( t-x \right)^{1-\alpha}}dx = \frac{1}{\Gamma(\alpha)}\int_{0}^{t} \frac{f(x)}{t^{1-\alpha}\left( 1-\frac{x}{t} \right)^{1-\alpha}}dx
\label{eq9}
\end{equation}

Setting $\frac{x}{t}=u$, with $dx=tdu$,

\begin{equation}
J^\alpha f(x) = \frac{t^\alpha}{\Gamma(\alpha)}\int_{0}^{1} \frac{f(tu)}{\left( 1-u \right)^{1-\alpha}}du
\end{equation}

To obtain the same interval of integration required in the Gauss-Mehler quadrature it is necessary to define $u=\frac{y+1}{2}$, that is, $du=\frac{dy}{2}$. Therefore:

\begin{equation}
J^{\alpha} f(x) = \frac{t^\alpha}{2^\alpha\Gamma(\alpha)}\int_{-1}^{1} \frac{f\left(\frac{t(y+1)}{2}\right)}{\left( 1 - y \right)^{1-\alpha}}dy
=
\frac{t^\alpha}{2^\alpha\Gamma(\alpha)}\int_{-1}^{1} \frac{h(y)}{(1-y^2)^\frac{1}{2}}dy
\end{equation}
%
%
%
with $h(y)=\frac{f((y+1)t/2)\left( 1-y^2 \right)^\frac{1}{2}}{\left( 1 - y \right)^{1-\alpha}}$.
Using this result in equation (\ref{eq8}),

\begin{equation}\label{eq:gmq:jalpha}
J^\alpha f(x) = \frac{t^\alpha}{2^\alpha\Gamma(\alpha)}\frac{\pi}{n}\sum_{i=1}^{n} \frac{f\left(\frac{(k+1)t}{2}\right)\left( 1 - k^2 \right)^\frac{1}{2}}{\left( 1 - k \right)^{1-\alpha}}
\end{equation}

which is the desired expression. Solving an integral of fractional order or not, can be easily performed with this equation. To validate this equation consider the function $f(t)=t^{\mu}$ for $t>0$ and $\mu>-1$, which has the exact solution for fractional integral:

\begin{equation}
J^{\alpha} f(t) = x^{\alpha+\mu}\frac{\Gamma\left(\mu+1\right)}{\Gamma\left(\mu+\alpha+1\right)}
\end{equation}

The calculated and exact results for $\alpha=0.5$, $\mu=1$ and $n=11$ are represented in Table \ref{table::exact:calculated}, showing the series solution can give almost exact result.\\

\begin{center}
	Table 1 goes here
\end{center}

Therefore, equation (\ref{eq:gmq:jalpha}) gives an excellent approximation for this example. It is important to emphasize that for each problem the number of points to a given precision should be analyzed.\\

The present approach indicates a way to generalize the fractional derivative using, 
	
\begin{equation}
D^\alpha f(x) = \frac{d^m}{dt^m}J^{m-\alpha} f(x) =\frac{d^m}{dt^m} \left[\frac{t^{m-\alpha}}{2^{m-\alpha}\Gamma(m-\alpha)}\frac{\pi}{n}\sum_{i=1}^{n} \frac{f\left(\frac{(k+1)t}{2}\right)\left( 1 - k^2 \right)^\frac{1}{2}}{\left( 1 - k \right)^{1-m+\alpha}}\right]
\end{equation}
For $0<\alpha<1$ one has,

\begin{equation}
\begin{gathered}
D^\alpha f(x) = \frac{\left(1-\alpha\right)t^{-\alpha}}{2^{1-\alpha}\Gamma(1-\alpha)}\frac{\pi}{n}\sum_{i=1}^{n} \frac{f\left(\frac{(k+1)t}{2}\right)\left( 1 - k^2 \right)^\frac{1}{2}}{\left( 1 - k \right)^{\alpha}} + \\
\frac{t^{1-\alpha}}{2^{1-\alpha}\Gamma(1-\alpha)}\frac{\pi}{n}\sum_{i=1}^{n}\frac{(k+1)}{2} \frac{f'\left(\frac{(k+1)t}{2}\right)\left( 1 - k^2 \right)^\frac{1}{2}}{\left( 1 - k \right)^{\alpha}}
\end{gathered}
\end{equation}
which is straighforward to apply, since it involves a simple summation. This can be carried out to any derivative, fractional order or not. \\

For the specific problem to represent the angle, equation (\ref{eq:gmq:jalpha}) will be tranformed into,

\begin{equation}
J^\frac{1}{2} \theta(E) = \frac{U^{\frac{1}{2}}}{2^{\frac{1}{2}}\Gamma\left(\frac{1}{2}\right)}\frac{\pi}{n}\sum_{i=1}^{n} \frac{\theta \left(\frac{(k+1)U}{2}\right)\left( 1 - k^2 \right)^\frac{1}{2}}{\left( 1 - k \right)^{\frac{1}{2}}}
\end{equation}
which will be applied to invert the angle. Therefore, the general form for $x(U)$ is given by

\begin{equation}\label{eq:xU:geral}
x(U)=\frac{1}{nL}\sqrt{\frac{2\mu U}{2}}\sum_{i=1}^{n} \theta \left(\frac{(k+1)U}{2}\right)\left( 1 + k \right)^\frac{1}{2}
\end{equation}
which can be used for both analytical form of $\theta(E)$ or interpolated results, as shown in the two examples discussed further.

\section{A first example: the Rutherford scattering}
The first problem to be inverted is the classical Rutherford scattering. Although this problem has an analytical solution it will be important to introduce the framework in which the inverse problem is to be carried out. More complicated potential will be treated next. In the original work, Rutherford obtained the Coulomb potential from the scattering angle data, using the physical and geometrical considerations of the problem. 
Rutherford deflection function is given by \cite{landau}

\begin{equation}\label{eq:theta:E}
\theta(E) = \frac{\pi}{2} - \cot^{-1}\left( \frac{L\sqrt{E}}{\alpha} \right)
\end{equation}
with 
$
\alpha = \sqrt{2\mu}\left( \frac{q_1q_2}{8\pi\epsilon_0} \right)
$, 
$q_i$ the particle charge and $\epsilon_0$ the vacuum permittivity. 
%
%
%
%
The collision between two electrons, $\mu=\frac{1}{2}$, with $b$ varying with $E$ to ensure $L$ constant, all in atomic units, will be considered. 
Equation (\ref{eq:theta:E}) was used to calculate the solution of the fractional integral (\ref{eq:xU}), using the result given by (\ref{eq:gmq:jalpha}). \\

%
%
%
%

Considering $n=4$, $L=0.5$ and $\mu=0.5$ in (\ref{eq:xU:geral}), one obtains:

%


\begin{equation}
x(U)=\frac{1}{2}\sqrt{\frac{U}{2}} \sum_{i=1}^{4} \theta \left(\frac{(k+1)U}{2}\right)\left( 1 + k \right)^\frac{1}{2}
\end{equation}
or,

\begin{equation}\label{eq:jmeioseries}
\begin{array}{lll}
x(U) & = & \frac{1}{2}\sqrt{\frac{U}{2}}\left[ \left( \frac{\pi}{2} - \cot^{-1}\left( \sqrt{\frac{1.9239U}{2}} \right) \right)1.3870 + \right.\\
&&\left( \frac{\pi}{2} - \cot^{-1}\left( \sqrt{\frac{1.3827U}{2}} \right) \right)1.1759 + \\
&& \left( \frac{\pi}{2} - \cot^{-1}\left( \sqrt{\frac{0.6173U}{2}} \right) \right)0.7857 +  \\
&&\left. \left( \frac{\pi}{2} - \cot^{-1}\left( \sqrt{\frac{0.0761U}{2}} \right) \right)0.2759 \right]
\end{array}
\end{equation}
%

Once the value of $x(U)$ is obtained, for different values of $U$, one can retrieve the potential energy function. The absolute relative error for this points are negligible and results for exact and calculated data are given in Table \ref{table:1}

\begin{center}
	Table 2 goes here
\end{center}

In this section it was observed that a four terms series is sufficient to retrieve the Coulomb potential very accurately.

\section{Lennard-Jones potential energy function}
Inverse procedure with fractional derivative has been applied, so far, to a Coulomb potential function, which has only the repulsive interaction. However, the methodology presented in this work is valid for more complex potentials with repulsive and atractive regions, such as a Lennard-Jones type potential. The general strategy for any system is illustrated as, \\


\begin{center}
	\framebox{Arbitrary values of $U$ are chosen}
\end{center}
\begin{center}
	$\downarrow$
\end{center}
\begin{center}
	\framebox{The parameters $n$, $\mu$ and $L$ are defined}
\end{center}
\begin{center}
	$\downarrow$
\end{center}
\begin{center}
	\framebox{The analytical or interpolated values for $\theta\left(\frac{(k+1)U}{2}\right)$ are obtained}
\end{center}
\begin{center}
	$\downarrow$
\end{center}
\begin{center}
	\framebox{These values are introduced in equation (\ref{eq:xU:geral})}
\end{center}
\begin{center}
	$\downarrow$
\end{center}
\begin{center}
	\framebox{$E_p(x) = U(x) - Eb^2x(U)^2$ with $x(U)=\frac{1}{r}$}
\end{center}
%

\begin{figure}[h!]\centering
	\includegraphics[width=13cm]{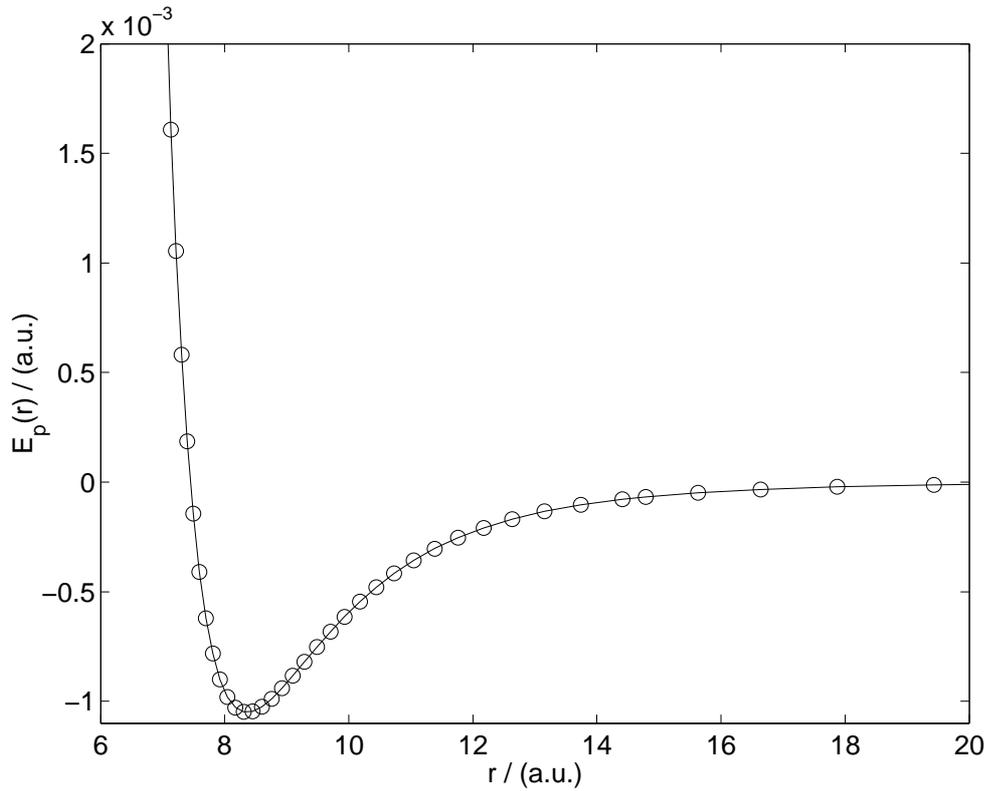}
	\caption{Lennard-Jones potential obtained by deflection angle and exact values plotted for data in the well region for $L=2500$.}
	\label{Fig:LJ:pot}
\end{figure}

The scattering between Xe and Rn was taken as a reference system, with simulated data. Scattering angle by energy were obtained solving the integral (\ref{int:theta}) with the impact parameter for each energy defined to maintain the angular momentum constant. The data calculated was interpolated and used in equation (\ref{eq:gmq:jalpha}).
Convergence with five significative was achieved with $n=407$ points in the quadrature for $x(U)$. 
For $L=bp$, energy and impact were varied to maintain the angular momentum constant, since one needs hyperbolic trajectory on the $bp$ plane. This step was not necessary for the Coulomb scattering, since analytical solution were available which guarantees the angular momentum as a constant of motion. Data were generated for $L=2500$, using 
$L=b\sqrt{2\mu E} \approx 548.43b\sqrt{E}$.
Inverted results
are plotted in Fig. \ref{Fig:LJ:pot}. All results are within 1.4\% of error, showing the applicability of the present algorithm. \\
%
%
%

\begin{center}
	Figure 1 goes here
\end{center}

For a potential energy surface with short and long range, such as a Lennard-Jones type potential, it was observed the need of more terms to retrieve accurate results. 
Even in this case there are no requirement of a great computational effort.

\section{Conclusion}
The overlap between the classical scattering theory, fractional calculus and inverse problems was presented in this work. The approach is an alternative way to the Firsov method or to the Abel classical solution to the problem.
The basic definitions to the fractional calculus development were given to provide a necessary theoretical background.
This theory was used to relate the classical scattering integral with the Riemann-Liouville half-order integral.
Changes of variables in the fractional integral were made and it was rewritten as a Gauss-Mehler quadrature and a simple example was given to validate the methodology.\\

Coulomb scattering, which has an analytical solution for the scattering angle, was used as a first test for the inversion procedure. A four term expansion in the integral was sufficient to achieve an excellent convergence for the potential, with negligible error. \\

The Lennard-Jones potential energy function, a more complex potential, was obtained by the same procedure. However, it was necessary a large number of points to obtain an acceptable accuracy. The data used in the inversion procedure were calculated theoretically, it was possible the retrieve the short and long range part of the potential. If experimental data were available the adopted strategy will be the same. This experimental data is possible by taking the differential cross for angles greater than the rainbow angle.\\


Therefore, the general scheme presented here gives an excellent accuracy, which can be increased even more by using more terms in the series expansion, regardless the potential energy function form. The only difference from a problem to another will be the number of points necessary to perform the inversion. This work opens the possibility to investigate the inverse quantum scattering process from the fractional calculus point of view.

\section{Acknowledgment}	
We would like to thank CNPq for financial support and to Professor Esequia Sauter at UFRGS for the helpful sugestions.
\clearpage


\clearpage

\begin{table}[h!]\centering
	\caption{Fractional derivative for $f(t)=t^{\mu}$ with $n=11$.}\label{table::exact:calculated}
	\begin{tabular}{ccc}
		\hline
		$t$&Calculated         & Exact \\
		\hline
		0&0                  & 0                  \\
		2&2.12768  & 2.12769   \\
		4&6.01799  & 6.01802   \\
		6&11.05576 & 11.05581  \\
		8&17.02146 & 17.02154  \\
		10&23.78821 & 23.78832 \\
		\hline
	\end{tabular}
\end{table}

\begin{table}[h!]\centering
	\caption{\label{table:1} Exact and calculated results for Coulomb potential energy function.}
	\begin{tabular}{cc}
		\hline
		$E_{p,cal}$ & $E_{p,exa}$\\
		\hline
		1.20624 & 1.20624 \\
		8.77499(-1) & 8.77499(-1) \\
		3.96078(-2) & 3.96078(-2) \\
		1.99010(-2) & 1.99010(-2) \\
		\hline
	\end{tabular}
\end{table}

\end{document}